\documentclass[preprint,10pt]{elsarticle}
\usepackage[utf8]{inputenc}

\usepackage[hyphens]{url}
\usepackage{listings}
\usepackage{xcolor}
\usepackage{graphicx}
\usepackage{subfig}
\usepackage{caption}
\usepackage{float}
\usepackage{marginnote}
\usepackage{comment}

\usepackage{hyperref}
\usepackage{color}

\usepackage{soul}

\definecolor{mGreen}{rgb}{0,0.6,0}
\definecolor{mGray}{rgb}{0.5,0.5,0.5}
\definecolor{mPurple}{rgb}{0.58,0,0.82}
\definecolor{backgroundColour}{rgb}{0.95,0.95,0.92}

\lstdefinestyle{CStyle}{
	backgroundcolor=\color{backgroundColour},   
	commentstyle=\color{mGreen},
	keywordstyle=\color{magenta},
	numberstyle=\tiny\color{mGray},
	stringstyle=\color{mPurple},
	basicstyle=\footnotesize,
	breakatwhitespace=false,         
	breaklines=true,                 
	captionpos=b,                    
	keepspaces=true,                 
	numbers=left,                    
	numbersep=5pt,                  
	showspaces=false,                
	showstringspaces=false,
	showtabs=false,                  
	tabsize=2,
	language=C++
}

\title{
PenRed: An extensible and parallel Monte-Carlo framework for radiation transport based on PENELOPE}
\author[i3m]{V.  Gim\'enez-Alventosa\corref{correspondingauthor}}
\ead{vicent.gimenez@i3m.upv.es}

\address[i3m]{Instituto de Instrumentaci\'on para Imagen Molecular (I3M)\\
Centro mixto CSIC - Universitat Polit\`ecnica de Val\`encia \\
Camí de Vera s/n, 46022, Val\`encia, Spain
} 

\author[uv]{V.  Gim\'enez G\'omez\corref{correspondingauthor}}
\ead{vicente.gimenez@uv.es}

\address[uv]{Departament de F\'{\i}sica Te\`orica and IFIC\\
Universitat de Val\`encia-CSIC \\
Dr. Moliner, 50, 46100, Burjassot, València, Spain
} 

\author[ISIRIYM]{S. Oliver}
\ead{sanolgi@upvnet.upv.es}

\address[ISIRIYM]{Instituto de Seguridad Industrial, Radiof\'isica y Medioambiental (ISIRYM) \\
Universitat Polit\`ecnica de Val\`encia\\ Camí de Vera s/n, 46022, Val\`encia, Spain} 

\begin{document}

\begin{abstract}
    
    Monte Carlo methods provide detailed and accurate results for radiation transport simulations. Unfortunately, the high computational cost of these methods limits its usage in real-time applications. Moreover, existing computer codes do not provide a methodology for adapting these kind of simulations to specific problems without advanced knowledge of the corresponding code system, and this restricts their applicability. To help solve these current limitations, we present PenRed, a general-purpose, stand-alone, extensible and modular framework code based on PENELOPE for parallel Monte Carlo simulations of electron-photon transport through matter. It has been implemented in C++ programming language and takes advantage of modern object-oriented technologies. In addition, PenRed offers the capability to read and process DICOM images as well as to construct and simulate image-based voxelized geometries, so as to facilitate its usage in medical applications. Our framework has been successfully verified against the original PENELOPE Fortran code. Furthermore, the implemented parallelism has been tested showing a significant improvement in the simulation time without any loss in precision of results.
    
\end{abstract}


\maketitle

\section{PROGRAM SUMMARY}

\begin{itemize}
    \item[] {\it Program title:} PenRed: Parallel Engine for Radiation Energy Deposition.
    \item[] {\it Licensing provision:} GNU Affero General Public License (AGPL).
    \item[] {\it Programming language:} C++ standard 2011.
    \item[] {\it Nature of problem:} Monte Carlo simulations usually require a huge amount of computation time to achieve low statistical uncertainties. In addition, many applications necessitate particular characteristics or the extraction of specific quantities from the simulation. However, most available Monte Carlo codes do not provide an efficient parallel and truly modular structure which allows users to easily customise their code to suit their needs without an in-depth knowledge of the code system.
    \item [] {\it Solution method:} PenRed is a fully parallel, modular and customizable framework for Monte Carlo simulations of the passage of radiation through matter. It is based on the PENELOPE [1] code system, from which inherits its unique physics models and tracking algorithms for charged particles. PenRed has been coded in C++ following an object-oriented programming paradigm restricted to the C++11 standard. Our engine implements parallelism via a double approach: on the one hand, by using standard C++ threads for shared memory, improving the access and usage of the memory, and, on the other hand, via the MPI standard for distributed memory infrastructures. Notice that both kinds of parallelism can be combined together in the same simulation. Moreover, both threads and MPI processes, can be balanced using the builtin load balance system (RUPER-LB [2]) to maximise the performance on heterogeneous infrastructures. In addition, PenRed provides a modular structure with methods designed to easily extend its functionality. Thus, users can create their own independent modules to adapt our engine to their needs without changing the original modules. Furthermore, user extensions will take advantage of the builtin parallelism without any extra effort or knowledge of parallel programming. 
    
    \item [] {\it Additional comments including Restrictions and Unusual features:}
    PenRed has been compiled in linux systems with {\tt g++} of GCC versions 4.8.5, 7.3.1, 8.3.1 and 9; clang version 3.4.2 and intel C++ compiler ({\tt icc}) version 19.0.5.281. Since it is a C++11-standard compliant code, PenRed should be able to compile with any compiler with C++11 support. In addition, if the code is compiled without MPI support, it does not require any non standard library. To enable MPI capabilities, the user needs to install whatever available MPI implementation, such as openMPI [3] or mpich [4], which can be found in the repositories of any linux distribution. Finally, to provide DICOM processing support, PenRed can be optionally compiled using the dicom toolkit (dcmtk) [5] library. Thus, PenRed has only two optional dependencies, an MPI implementation and the dcmtk library.
    
    \item [] {\it References:\\}
    [1] F.  Salvat, penelope-2018: A code System for Monte Carlo Simulation of Electron and Photon Transport, OECD/NEA Data Bank, Issy-les-Moulineaux, France, 2019, available from http://www.nea.fr/lists/pene\-lope.html
    
    [2] V. G. Alventosa, G. M. Martínez, J. D. S. Quilis, RUPER-LB: Load balancing embarrasingly  parallel  applications  in  unpredictable  cloud  environments (2020). arXiv:2005.06361.
    
    [3] Graham R.L., Woodall T.S., Squyres J.M. (2006) Open MPI: A Flexible High Performance MPI. In: Wyrzykowski R., Dongarra J., Meyer N., Waśniewski J. (eds) Parallel Processing and Applied Mathematics. PPAM 2005. Lecture Notes in Computer Science, vol 3911. Springer, Berlin, Heidelberg

    [4] William Gropp. 2002. MPICH2: A New Start for MPI Implementations. In Proceedings of the 9th European PVM/MPI Users’ Group Meeting on Recent Advances in Parallel Virtual Machine and Message Passing Interface. Springer-Verlag, Berlin, Heidelberg, 7.
    
    [5] DCMTK, https://github.com/DCMTK/dcmtk, accessed: 2019-09-28
    
\end{itemize}

\section{Introduction}
Monte-Carlo (MC) methods are widely used in most scientific applications which involve radiation transport simulations, including electron microscopy and microanalysis, x-ray fluorescence, detector characterisation, radiation me\-tro\-logy, dosimetry and radiotherapy, among others.

MC are statistically based methods that, among many other applications, can be used to resolve radiation transport problems by random sampling. Detailed chronological MC simulations yield the same results as the solution of the linear Boltzmann transport equation, within statistical error bars. Thus, the Type A uncertainties associated to MC simulations are strongly dependent on the number of sampled particle histories. On the other hand, Type B uncertainties may occur because of the usage of variance reduction techniques, such as condensed-history algorithms, particle splitting etc. Moreover, the Type A uncertainties strongly depend on the complexity of the system geometry, and the distance to the source, due to the decreasing of particle fluence, and other configuration parameters. However, even if the statistical uncertainties can be reduced by increasing the number of histories and the calculation time, they decrease only as $1/\sqrt{N}$ \cite{Salvat2019}, what means that, for instance, it is required a statistics $100$ times larger to reduce the uncertainty by a factor of $10$. Usually, particle transport codes based on MC methods scale their execution time linearly with the number of histories, therefore, in the previous example, the calculation time will be increased by a factor of $100$.

Another point to take into account in MC simulation codes is their usage field. The huge amount of possible applications and configurations of the simulation system (energy ranges, geometries, particle types and physics models, etc.) make difficult, if not impossible, to cover all options on a single MC code system. As a consequence, there exist different MC code system packages as EGS \cite{egs}, MCNP \cite{MCNP}, PENELOPE \cite{PENELOPE}, GEANT IV \cite{GEANT4} or FLUKA \cite{FLUKA} and many adaptations for specific applications. Among them, PENELOPE stands out for its accurate implementation of electron and positron electromagnetic physics, has an open source license and its source code is relatively simple compared with other MC codes. Furthermore, several previous works have parallelized PENELOPE to deal with the long execution times \cite{BADAL2006440}, some of them focused on medical applications using MPI parallelism \cite{Cruise2003PARALLELIZATIONOT} or GPU acceleration \cite{doi:10.1118/1.3231824}. Some works also use simplifications to accelerate the simulations for specific purposes \cite{Rodrguez2018DPMAA}.

However, the existent parallelizations and optimizations of the PENELOPE code have some disadvantages, such as loss of generality, and, for instance, limited applicability beyond its specific purpose, or low capability for running efficiently on heterogeneous architectures. These limitations will be discussed in detail in section \ref{sec:stateArt}.

The present work is aimed to provide a general purpose, highly parallel, efficient and flexible MC simulation framework, called PenRed  (Parallel  ENgine  for Radiation Energy Deposition). It contains a restructured object oriented (OO) C++ translation of the original PENELOPE library to provide a modular, parallel and easily extensible code system. PenRed is distributed as a free and open source program and can be downloaded from its repository\footnote{\url{https://github.com/PenRed/PenRed}}\footnote{\url{https://archive.softwareheritage.org/swh:1:dir:87fd3a7e44d76653486135914430a87fd60e92ac/}}.

The rest of the paper is organized as follows. In Section 3, a brief revision of the state of the art is presented, and the advantages of PenRed versus already existing PENELOPE based codes are discussed. A description of both the translation and restructuring of the code, the PenRed capability to be easily customised, as well as its parallelism model and implementation are discussed in Section 4. 
In Section 5, the validation of our code is described in detail. Section 6, contains a comparison of the performance of parallel executions among PenRed, PENELOPE and PenEasy. In addition, it includes a study of the PenRed behaviour on parallel executions. Finally, current research and future plans of development are described in Section 7.

\section{State of the art}
\label{sec:stateArt}
Our revision of the state of the art of the optimisation of PENELOPE is based on the study of the works \cite{BADAL2006440, Cruise2003PARALLELIZATIONOT, doi:10.1118/1.3231824, Rodrguez2018DPMAA}. 

There are some works that optimised the code for a specific application, targeting to increase speed but compromising its range of application. One of these works is the DPM code \cite{Rodrguez2018DPMAA}, which uses some approximations of the particle transport description and the underlying physics models, for instance, simplified cross sections, to reduce the simulation run-time. The disadvantage of this approach is that it is accurate only for low atomic numbers and in a specific energy range, typically the one used in conventional radiotherapy treatments. Because of its high efficiency, other works have been focused on the optimisation of DPM, some using Message Passing Interface (MPI) \cite{doi:10.1118/1.1786691}, and others vectorizing the code \cite{Weng_2003}, or adapting it to Graphics Processing Units (GPU) \cite{Jia_2010}\cite{Wang_2017}.

On the other hand, \cite{Cruise2003PARALLELIZATIONOT} presents an MPI version of PENELOPE. Although authors are focused on medical applications, and the tests performed belong to this field, the code should run generic simulations using the PENELOPE library. The code has been applied to perform simulations of photon beams using a method for treating intracranial lesions. However, this work does not provide a multithreading implementation, which could take advantage of sharing resources and reduce the memory usage. In addition, a modular and extensible structure based on OO programming is unavailable.

Let us turn now to the GPU acceleration of PENELOPE. In \cite{doi:10.1118/1.3231824} the authors develop with the CUDA programming model (NVIDIA Corporation, Santa Clara, CA), a MC simulation code which runs entirely on a GPU using the photon interaction model of the PENELOPE code. Their approach focuses on the simulation of x-ray transport, thus neither electron nor positron transport is considered. Through the simulation of voxel based geometries, and simplifications such as the usage of single precision operations, the authors achieve a speed one order of magnitude faster than the CPU version. Although their GPU computing is very efficient for their purpose, it is limited to specific applications. A similar approach is used by \cite{doi:10.1118/1.4953198}, which develops a GPU-accelerated MC dose calculation platform based on PENELOPE focused on commissioning of IMRT. Moreover, the authors ensure that their approach should be applicable to nearly any application requiring high dose accuracy.

Finally, \cite{BADAL2006440} presents a package of Linux scripts for the parallelization of MC simulations with PENELOPE. Just as \cite{Cruise2003PARALLELIZATIONOT}, this work uses PENELOPE with no approximations or limitations. However the same disadvantages apply to this study. Furthermore, these works do not provide mechanisms to take full advantage of heterogeneous systems, where some kind of load balancing is required.

In addition to optimisation based works, PenEasy \cite{sempau2006peneasy} was developed as an easy and modular main program for PENELOPE. However, the former has not implemented any parallelism nor takes advantage of advanced OO modular programming.

A further shortcoming of most of the previous studies is that they use an outdated version of the PENELOPE library. An exception is PenEasy, which is updated frequently to use the latest PENELOPE version.

\section{Material and Methods}
\label{sec:MatiMet}

\begin{figure}[H]
    \centering
    \includegraphics[scale=0.5]{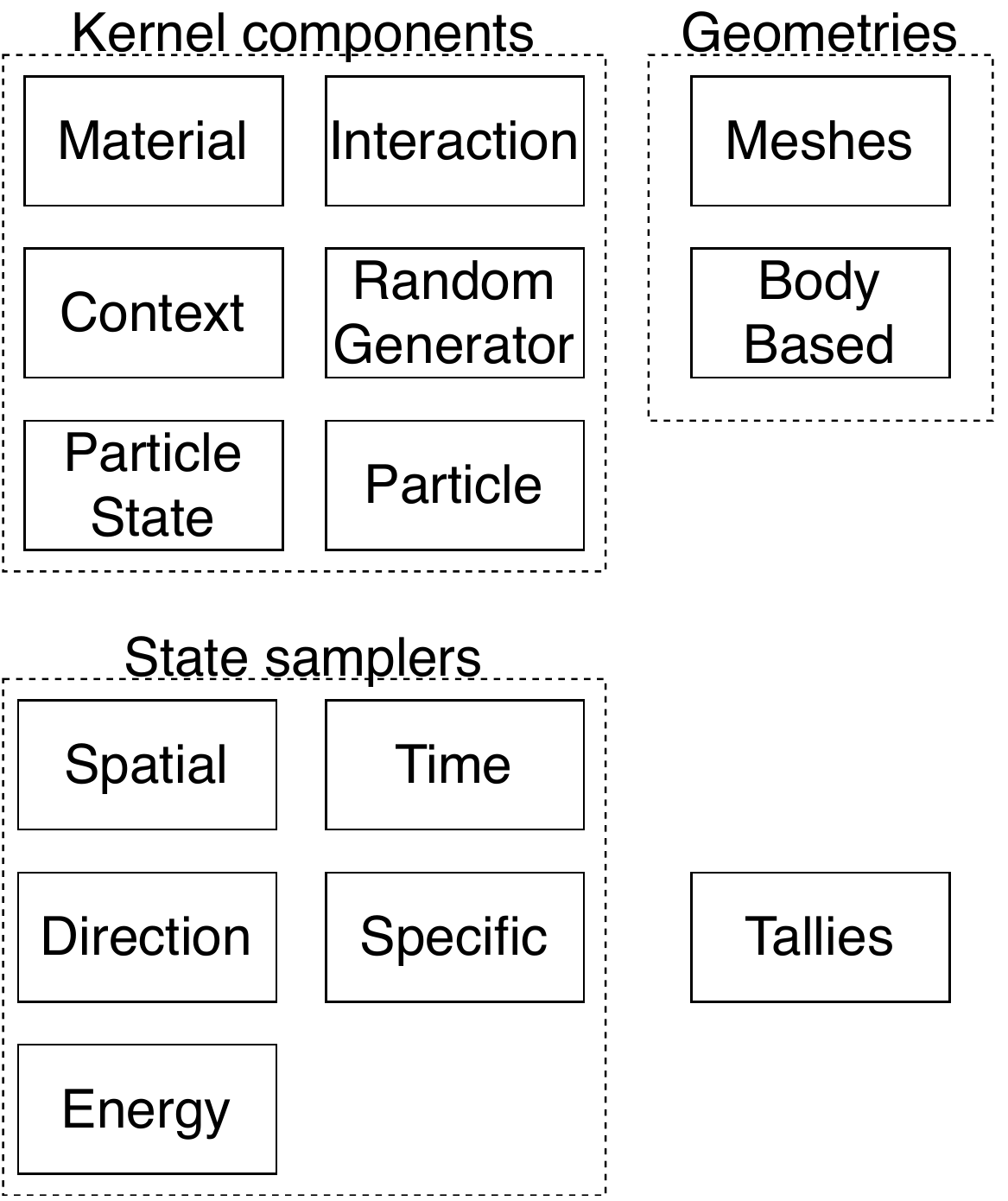}
    \caption{Classification of the PenRed modules.}
    \label{fig:components}
\end{figure}

PenRed has been organised using modules providing abstract template classes as interfaces to facilitate their customisation. These components, or modules, are shown in figure \ref{fig:components}, where PENELOPE whole library, including its geometry package, is implemented. A detailed description of each component and the implementation of the corresponding derived classes can be found in the PenRed documentation, which is distributed with the PenRed package. Briefly, {\tt kernel} components handle all the physics related tasks, the module {\tt geometries} takes care of particle transport across the geometry system, {\tt state samplers} determines the initial state of the particle and, finally, {\tt tallies} extracts information from the simulation loop.

Moreover, PenRed provides a main program, {\tt pen\_main}, to facilitate the usage of the package. In order to utilize it, the user only needs to specify the configuration of the simulation without coding anything.

\subsection{C++ translation}
\label{sec:transAndVerif}

The complete PENELOPE source code has been translated by hand to C++ and restructured using an OO approach to achieve a highly modular and extensible system. To this aim, some other codes with OO implementations make an intensive use of virtual methods. However, the recourse to virtual tables to perform calls to virtual methods involves a non negligible overhead \cite{10.1145/236337.236369}, specially on methods which are called continuously during the particle transport. To achieve a compromise between flexibility and performance, PenRed employs template abstract classes as interfaces which allows to use final derived classes on kernel components instead of a pointer to the abstract base class, avoiding the requirement to perform virtual calls in most cases. Of course, custom derived classes created by the user are intended to follow the same mechanism. During the code restructuring, an analysis to identify and optimise the most critical sections of the source code has been carried out. This analysis has been performed for all the examples included in the PENELOPE package, using the Valgrind \cite{valgrind} software with the Callgrind tool \cite{callgrind}, which provides a detailed information about the number of instructions required by each class, function and line of code on a simulated hardware. We found that, as expected, the most time-consuming functions for both languages (C++ and Fortran) are the particle simulation loops, and, within them, the {\it JUMP}, {\it STEP} and {\it KNOCK} subroutines. Furthermore, these three functions together represent about  $80-90\%$ of the simulation loop time in the profiled examples. As a consequence, several optimisations have been made specially in these functions, such as grouping specific variables and arrays into structures and structure arrays, respectively, to improve memory access.

\subsection{Extension capabilities}
\label{sec:flexibility}

As mentioned above, PenRed uses template abstract classes as interfaces to extend the code functionality by creating a derived class of the corresponding module. The method to extend a module is explained in the documentation. However, in this section we will focus only on the most useful for a standard user, namely, {\tt state samplers}, {\tt geometries} and {\tt tallies}.

\subsubsection{State samplers}
\label{sec:stateSamplers}

There are many {\tt state samplers} types, as shown in figure \ref{fig:components}. Among them, the spatial, time, direction and energy sampler types take care of determining the initial particle position, time of flight, direction and energy, respectively. These groups of samplers are named generic samplers, because they can be used for any particle regardless of the particle type. On the other hand, specific samplers can change the whole particle state but they are particle specific. They have been used, for example, to sample the Stokes parameters of polarised photons. 

This approach facilitates the creation of new samplers, since they can be coded as small classes using only two mandatory methods, the configuration and the sampling method itself. In addition, any sampler type can be combined with any other type, allowing creating complex state sampling configurations from simple parts.

Finally, new samplers can be easily added to the framework and they will be automatically accessible from the provided main program, without changing the code outside the new custom sampler class. The only required change is to add the corresponding source file names into two include files. Further details can be found in the PenRed documentation.

\subsubsection{Geometries}
\label{sec:geometries}

As shown in figure \ref{fig:components}, {\tt geometries} are classified into two main types, meshes and body based geometries. Although new geometry types can be created directly from the base abstract class, in order to facilitate the coding PenRed provides two specific interfaces to create geometries based on meshes and geometries based on bodies or geometric objects. At present, PenRed implements an optimised translation of the original PENELOPE geometry library as a body based geometry, and additionally, a voxelized geometry, from which the PenRed DICOM geometry has been created as a mesh based one. The voxelized geometry transport is based both on the penEasy voxelized transport \cite{sempau2006peneasy} and the PENELOPE penCT program \cite{SalvatPenCT}.

Using these intermediate interfaces, the creation of new geometries requires, mainly, the definition of the configuration and the {\it STEP} and {\it LOCATE} methods, i.e. the implementation of the geometry itself without taking care of framework details. The {\it STEP} and {\it LOCATE} methods handle, respectively, the transport of a particle across the geometry, and the localisation of a particle inside the geometry.

Regarding how to incorporate new geometries into PenRed and how to use them with the provided main program, the methodology is analogous to the one used by state samplers (see section \ref{sec:stateSamplers}).

\subsubsection{Tallies}
\label{sec:tallies}

\begin{figure}
    \centering
    \includegraphics[scale=0.41]{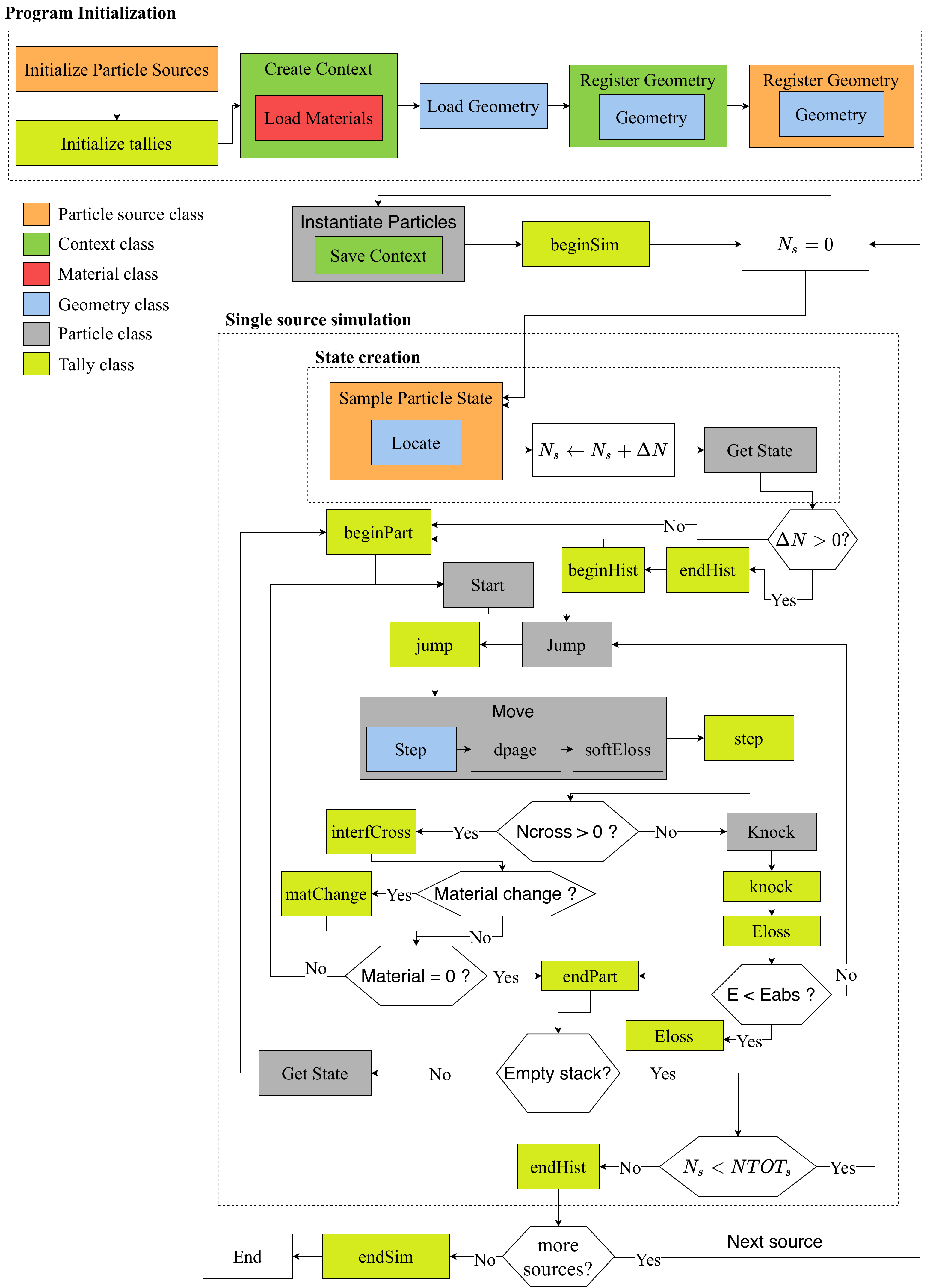}
    \caption{Basic flow diagram of the PenRed main program including tally calls.}
    \label{fig:flowTally}
\end{figure}

PenRed implements several tallies, most of them adapted from the main program of the original PENELOPE package. All of these are extensively described in the PenRed documentation. The capability to extend this component, i.e. to create new tallies, is, probably, the most valuable feature for a standard user. As other components, tallies are implemented via derived classes. However, to provide a generic interface that does not constrain the tally creation, their interface contains a number of virtual functions that will be called by the main program at different locations in the simulation loop. Notice that only the functions directly related to the information to be extracted by the tally, need to be implemented. Figure \ref{fig:flowTally} shows a simplified flow diagram where the tally functions (yellow boxes) are called at different points of the main loop. A complete description of the data registered by each function and when are called in the simulation loop can be found in the documentation. Again, the method to incorporate new tallies to the framework is the same as for samplers and geometries.

\subsubsection{Overhead}

To be able to include avoiding any recoding new implementations of the components described above, PenRed uses virtual calls. As discussed before, these calls could produce a non negligible run-time overhead. However, since the sampler process is usually executed once per history, the performance overhead generated by samplers is truly negligible. Moreover, the only virtual calls carried out by the geometry class are the {\it STEP} and the {\it LOCATE} functions. Although these two methods are called in every loop, they are only called at most once. In addition, they are generally computationally expensive, so the virtual call overhead is usually negligible as compared to the total function execution cost. Unfortunately, tally functions are called several times along the simulation loop, leading to a non negligible time overhead. In fact, we observe in some of the PENELOPE examples that up to $5\%$ of the total execution time has been wasted in virtual calls. 

Obviously, the number of virtual calls due to tally functions increases with the number of tallies invoked. Therefore, to reduce their footprint on critical applications, one solution is to create a wrapper tally containing many other tallies. With this approach only one virtual call per used function will be performed by the wrapper tally, while the wrapped tally functions can be called invoking the final class without accessing to the virtual table.

Note that most examples included in the PENELOPE package use a very simple geometry. As the complexity of the geometry grows, the influence of the {\tt STEP} function on the execution time grows too, minimising the relative weight of the virtual call overhead.

\subsection{Parallelism}
\label{sec:parallelism}

To take advantage of both, multi-core processor architectures and distributed infrastructures, PenRed implements parallelism at both levels, i.e. multithreading and multiprocessing. Notice that both parallelisms are optional and can be separately enabled or disabled during compilation. However, parallel coding is not a trivial task and may produce several outcomes that block our program (deadlocks), hide bugs, as race conditions, or effects that slow the whole program, like the false sharing effect \cite{Bolosky:1993:FSE:1295480.1295483}. Nevertheless, two undesirable approaches to combine high parallelism and extensibility are to limit the capability to extend the code according to the knowledge of the user on parallel programming or limit the parallelism itself. Instead, PenRed uses another approach consisting of providing a high grain parallelism completely transparent to the user. In this way, the user can write their modules like sequential code without worrying about parallelism and they will be able to be executed on multithreading and multiprocess runs.

To achieve this goal, specially on multithreading executions, PenRed limits the member functions that can change the state of the corresponding class. For example, geometry {\it STEP} and {\it LOCATE} member functions are defined as constant, so that they cannot change the state of the class, ensuring that the state of the geometry remains constant once initialised until the end of the simulation. An alternative approach could be to duplicate the whole simulation (geometry, samplers, etc). However, this approach generates a huge amount of duplicated memory, increasing its consume, and avoids threads to share memory at cache level, increasing the memory fails \cite{752659}.

The only module that requires a special,  but quite simple, treatment to be simulated on parallel executions are the {\tt tallies}. As PenRed cannot know how to add the partial results of a specific tally, a corresponding sum function must be implemented. This function gets an object of the same tally type as argument and it is supposed to sum both contributions and store the result on the object that calls the function. Thus, the only requirement consists of specifying how to sum the tally individual results. Other aspects as when the sum function is called or the order to reduce the results, are handled automatically by PenRed. The implemented tallies follow the guidelines described in the PENELOPE manual \cite{PENELOPE} to calculate the uncertainties of the scored magnitudes. That is, on sequential executions, the scored contributions of each history to a magnitude $q$ and its square $q^2$ are added to independent counters. Therefore, the estimators $Q$ and $Q^2$ of the magnitude and its square can be obtained as,

\begin{equation}
    \overline{Q} = \frac{1}{N} \sum^N_{i=1} q_i \;\;\;\;\;\; \overline{Q^2} = \frac{1}{N} \sum^N_{i=1} q^2_i
\end{equation}

\noindent where $N$ is the number of independent histories and $i$ indicates the history number. Then, the corresponding standard deviation is calculated as,

\begin{equation}
    \sigma_Q = \frac{1}{\sqrt{N}} \left[\overline{Q^2} - \overline{Q}^2 \right]
    \label{eq:seqVar}
\end{equation}

On the other hand, in multithreading executions, each thread uses independent sets of counters to sum the contributions of each history ($q_{ji}$, $q^2_{ji}$), where $i$ denotes the history index of thread $j$. Using these counters, the final estimators are,

\begin{equation}
    \overline{Q} = \frac{1}{N_t} \sum^{N_{th}}_{j=1} \sum^{N_{j}}_{i=1} q_{ji} \;\;\;\;\;\; \overline{Q^2} = \frac{1}{N_t} \sum^{N_{th}}_{j=1} \sum^{N_{j}}_{i=1} q^2_{ji}
    \label{eq:parallEstim}
\end{equation}

\noindent where $N_{t}$ is the total number of histories performed by all threads, $N_{th}$ is the number of threads, and $N_j$ is the number of histories simulated by the thread number $j$. At the end of the simulation, the partial results of all threads are summed and the average and standard deviation are calculated using eq.\ \ref{eq:parallEstim} and eq.\ \ref{eq:seqVar}, with $N\, =\, N_{t}$. The same methodology is used for multiprocessing and combined multithreading and multiprocess executions.

A very important issue to take into account on parallel executions is that we must ensure that each thread uses an independent succession of random numbers. Otherwise, different threads may produce correlated results. Since PenRed uses the same random number generator as the original PENELOPE code system, developed by F. James \cite{james}, independent sequences of random numbers are achieved using the seeds calculated by \cite{BADAL2006440}. These are automatically assigned by the PenRed main program to each thread on each process to avoid correlated results.

Concerning the libraries employed in PenRed for parallelism, multithreading has been implemented using the standard threads of C++11. Thus, it is not necessary any additional external library. On the other hand, multiprocess has been implemented via the MPI standard, hence any library that implements that standard, such as OpenMPI \cite{openMPI} or MPICH2 \cite{MPICH2}, can be used to run PenRed with MPI enabled.

Finally, it is important to notice that most modern distributed infrastructures are heterogeneous, i.e. each node has different capabilities, such as different processors, amount of memory, disk throughput and IOPS etc. Even if the system is homogeneous, the usage of an  infrastructure by many users could produce a non negligible overhead. For example, modern cloud computing infrastructures suffer unpredictable capability fluctuations \cite{5948601, 10.1145/2885497, 10.14778/1920841.1920902, doi:10.1080/02564602.2017.1393353}. To take full advantage of distributed infrastructures, PenRed incorporates a load balance system named RUPER-LB \cite{alventosa2020ruperlb} which automatically handles the assignation of histories to each thread in each process. Similarly to multithreading and MPI features, the load balance system can be enabled or disabled during the compilation process.

\subsection{DICOMs}
\label{sec:dicom}

As can be seen from most of the works reviewed in section \ref{sec:stateArt}, medical applications are a field of interest that requires huge optimisations for PENELOPE users. This is the reason why PenRed implements a special type of mesh based geometry to be able to perform simulations directly based on DICOM (Digital Imaging and Communications in Medicine) \cite{DICOM} images.

To offer support for this image format, PenRed uses the open-source library DICOM ToolKit (DCMTK) to extract all the required data from DICOM files. This library can be found in some linux repositories or github \cite{DCMTK_git}. As other features, DICOM support is optional, allowing to use PenRed without that library and its dependencies.

Currently, PenRed's DICOM module converts a CT or US DICOM image to a voxel geometry. To transform CT Hounsfield Units (HU) to density, the DICOM module requires a CT calibration curve to be provided by the user. Then, the material assignation could be done using the contours included in DICOM files or both by HU or density ranges. This approach allows users to perform the segmentation of CT images according to the guidelines of the TG-186 \cite{TG186}. On the other hand, US images require contour information to carry out the material and density assignations.

\section{Validation}
\label{sec:verification}

In this section, a collection of tests designed to verify the functionality of the PenRed framework are presented. The tests consist of a comparison between PENELOPE and PenRed of the sampling of the cross sections for all the interactions included in the original PENELOPE Fortran code. In addition, all the results of the examples distributed in the Penelope package are reproduced with and without parallelism. Finally, to validate the processing of voxel/DICOM geometries, a comparison with a GATE \cite{Jan_2011} DICOM simulation example is carried out.

For the sake of brevity, the results of the MPI tests will not be described here because their conclusions and figures are completely equivalent to those obtained from the multi-threading analysis (see section \ref{sec:multithreading}). Moreover, since MPI processes communicate with each other only in the post-processing step, and its contribution to the total simulation time is completely negligible on distributed memory infrastructures, it is useless to discuss a scalability analysis of MPI executions. In fact, the scalability, in simulations with negligible tally sum processing time, is determined by the slowest process and it is approximately ``perfect'' on a homogeneous cluster.

Most of the implemented features for quadric geometries can be validated by performing the simulations of all the PENELOPE examples. To reproduce our results, the user can find all the examples with the corresponding materials, geometries and configuration files, in the directory {\tt examples} of the PenRed distribution package.

As for the verification of voxel/DICOM geometries, basic tests consist of converting quadric geometries to voxel geometries and carrying out the very same simulation on both geometry types. The conversion can be performed through the PenRed provided utility {\tt geo2voxel},  which transforms any geometry into a voxelised one by means of the {\it LOCATE} method. The output file contains a voxel geometry ready to be simulated.
We have verified that the results from quadric and voxel geometries are perfectly compatible within statistical error bars. For the purpose of brevity, we will not discuss the results of these tests here. Our results however can be easily reproduced by the user using the tools provided in the package distribution. Instead, a more complete test consisting of a simulation on a DICOM image, is presented in section \ref{sec:DICOMtest}.

To run the tests, we used a single node with two Intel(R) Xeon(R) CPU E5-2660 v3 @ $2.60$ GHz processors, $8$ TB of disk storage and $125$ GB of memory RAM. Each of these processors had $10$ physical cores with hyperthreading, i.e. a total of $20$ physical and $40$ logical threads. The PenRed modules and the {\tt pen\_main} program were compiled using the {\tt g++} GNU C++ Compiler version 7.3.1 \cite{GCC}, on a Centos 7.0 Linux operating system.

\subsection{Cross-section sampling tests}
\label{sec:interactionTests}

To check the interaction sampling methods, all the differential cross sections (DCS) of each physics model implemented in PENELOPE have been sampled with PENELOPE and PenRed and their results compared. Notice that these tests employ the physics routines as isolated components, i.e. no other components such as the geometry package, tallies or a main program have been used. These tests have been repeated for different materials and energies. The conclusion is that all the sampled differential cross sections for each material, energy and interaction for both codes match exactly. These results ensure that the restructured physics modules implemented in PenRed are perfectly compatible with those of the original Fortran code.

\begin{figure}[htb]
\centering
\begin{subfloat}
    \centering
    \includegraphics[width=0.49\textwidth]{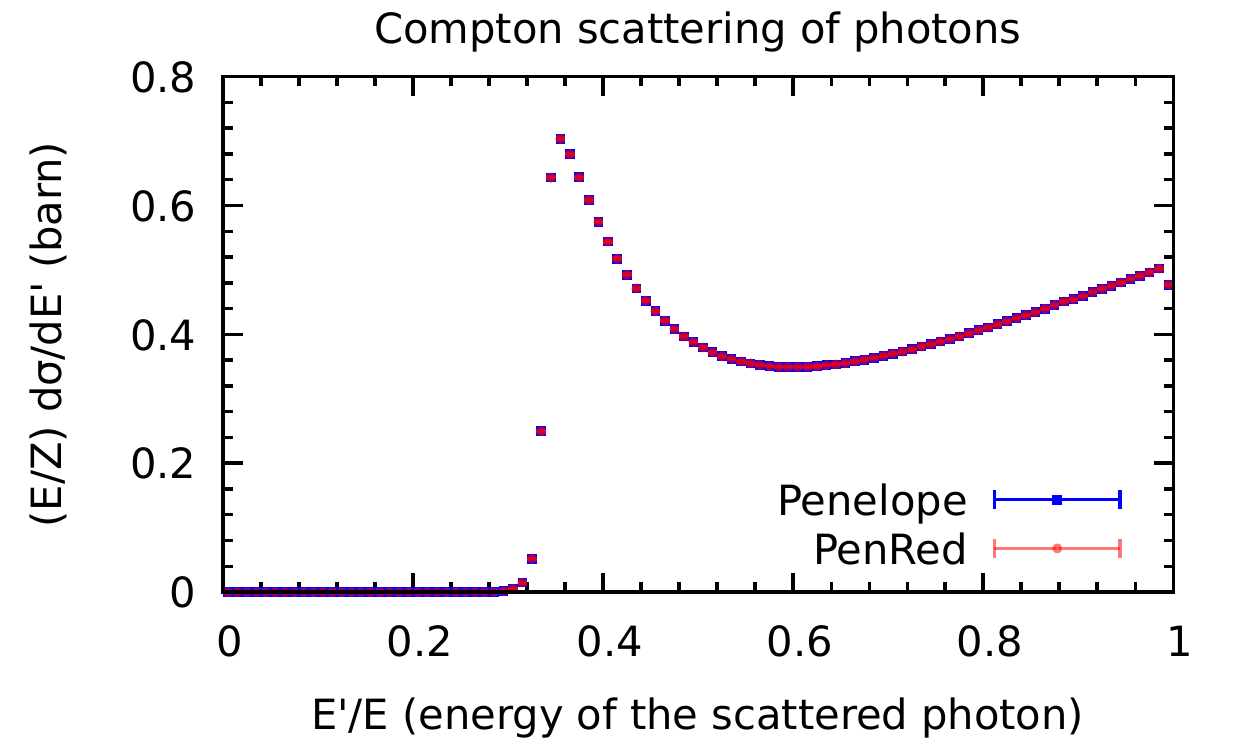}
\end{subfloat}
\hfill
\begin{subfloat}
    \centering
    \includegraphics[width=0.49\textwidth]{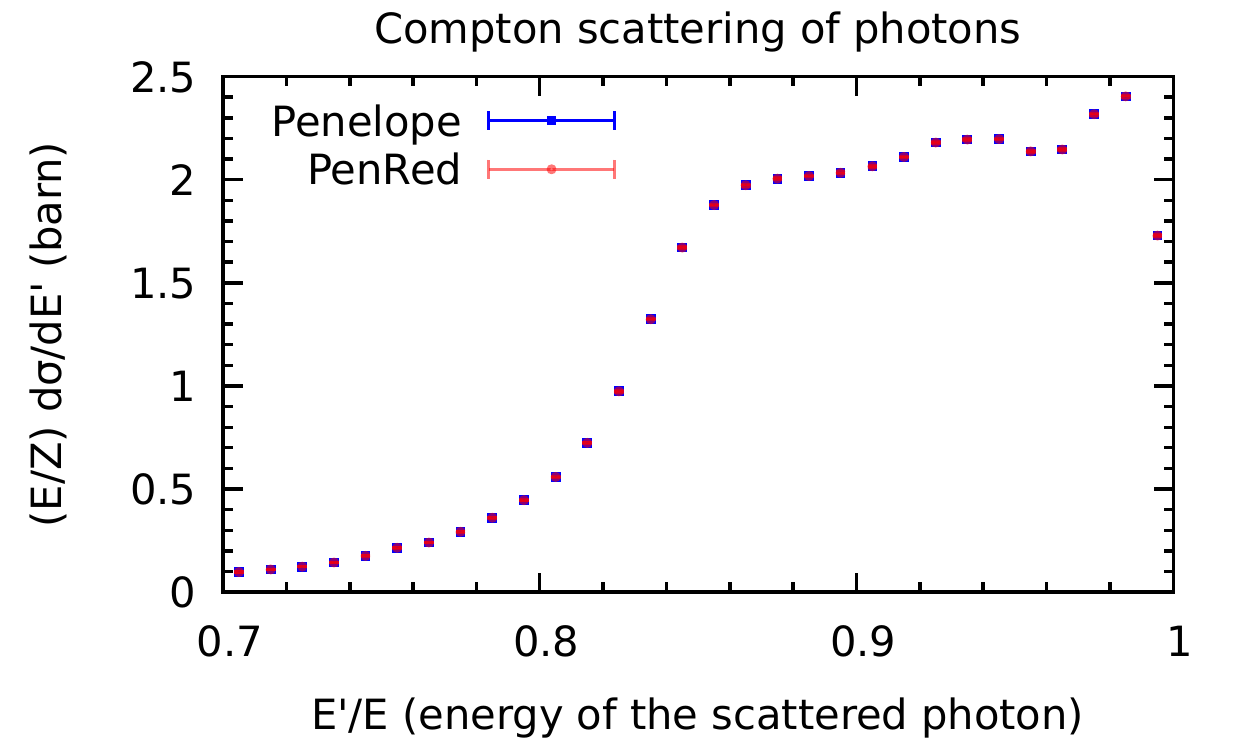}
\end{subfloat}
\vskip 0.5cm
\begin{subfloat}
    \centering
    \includegraphics[width=0.49\textwidth]{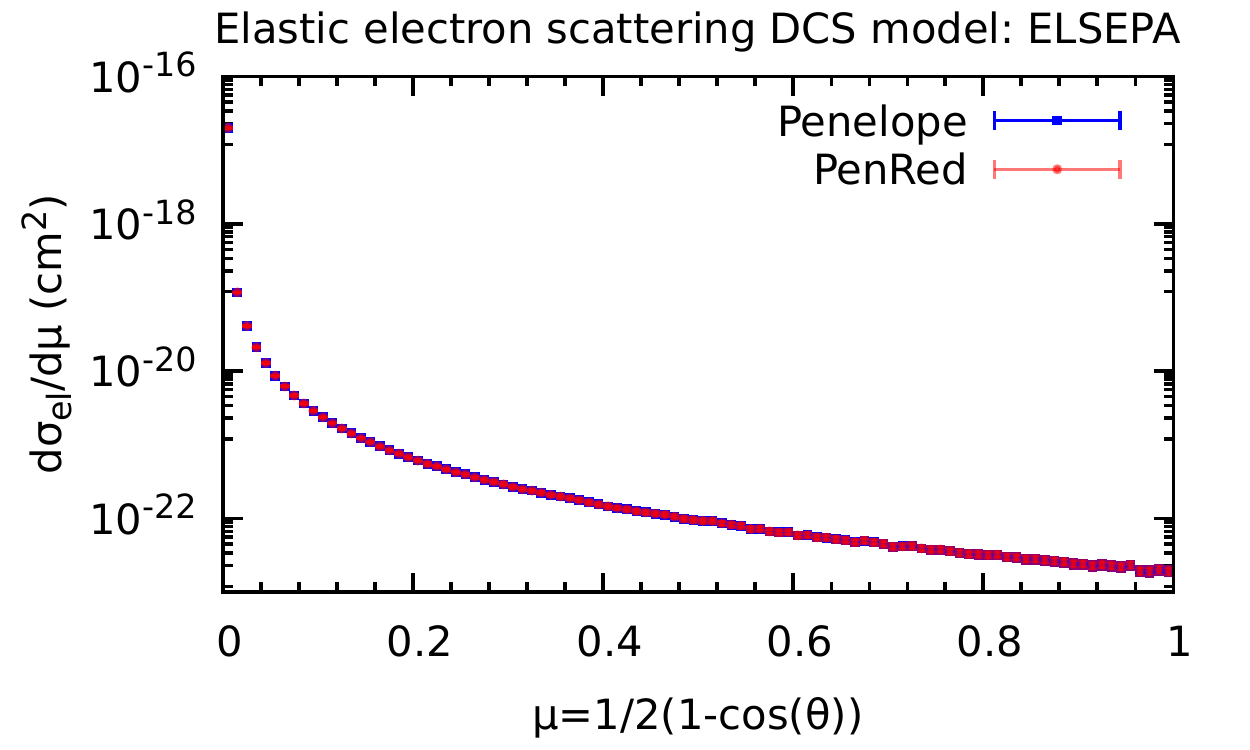}
\end{subfloat}
\hfill
\begin{subfloat}
    \centering
    \includegraphics[width=0.49\textwidth]{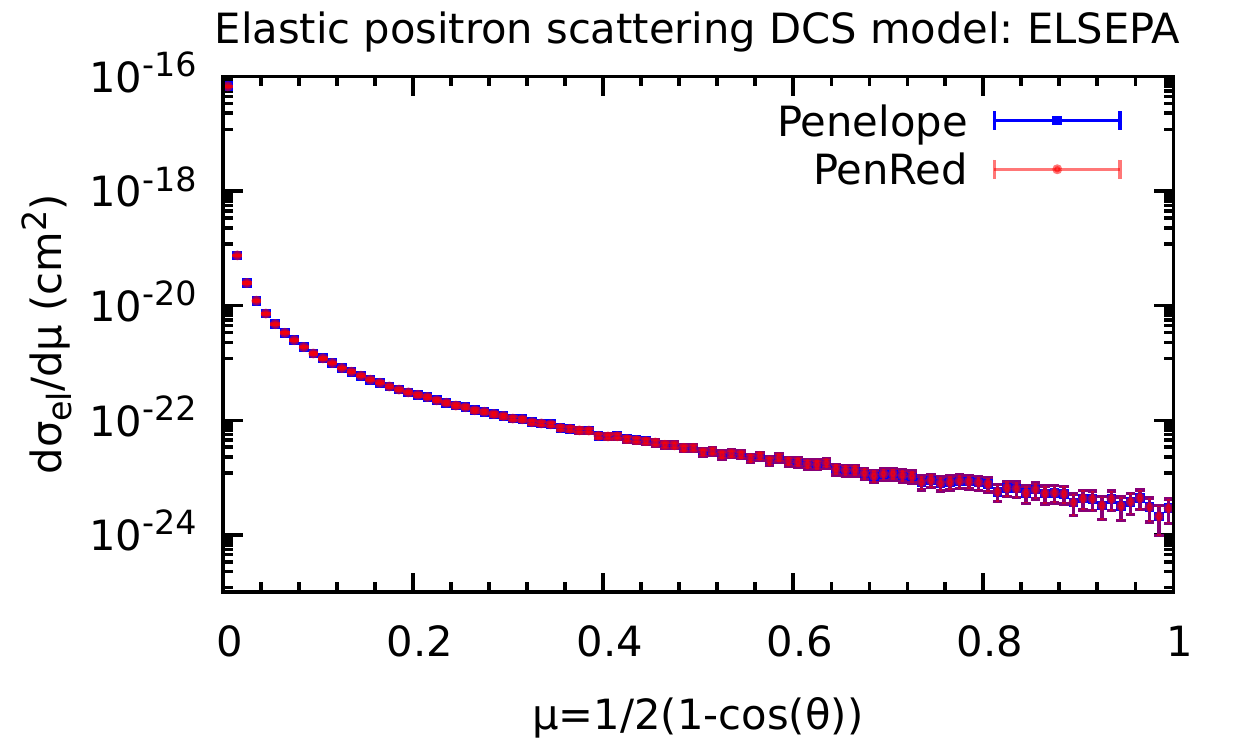}
\end{subfloat}
    \caption{Simulation results for the Compton energy differential cross section for scattered photons by aluminium ($Z=13$) with an incident photon energy of $500$ keV (upper left graph) and by gold ($Z=79$) with an incident energy of $50$ keV (upper right graph), and for the partial wave model DCS for elastic scattering of electrons by aluminium (lower left graph) and positrons by gold (lower right graph), with energies of $100$ keV and $1$ MeV, respectively. Error bars enclosure a deviation of $2\sigma$.}
    \label{fig:samplerGCO}
\end{figure}

Figure \ref{fig:samplerGCO} displays comparisons between the results from PENELOPE and PenRed for the Compton energy differential cross section for scattered photons sampled for aluminium (upper left graph) and gold (upper right graph) with incident photons of $500$ and $50$ keV, respectively, and for the partial wave model DCS for elastic scattering of electrons by aluminium (lower left  graph) and positrons by gold (lower right graph) with incident energies set to $100$ keV for electrons and $1$ MeV for positrons. As can be seen, the cross sections from PenRed are almost identical to those from PENELOPE.

\subsection{PENELOPE example tests}
\label{sec:PENELOPEtests}

In order to validate PenRed, the simulation results from all the setups of the examples included in the PENELOPE Fortran package were reproduced. These tests were done using a single thread. The comparison between PENELOPE and PenRed was carried out first by plotting all output files for the corresponding tallies, and second, to ensure that these codes were statistically compatible, by a bin-by-bin analysis of the histograms of the differences. Notice that the particle tracks generated by PENELOPE and PenRed follow in general different paths, in spite of the fact that both the random generator and initial seeds were the same. The reason is twofold: on the one hand, the differences in the main program structure, such as the use in PenRed of independent stacks for each particle type, and on the other hand, the different round-off errors that occur in the calculation of intersections of particle trajectories with material interfaces.
We have verified that the results from PENELOPE and PenRed are perfectly equivalent in all the cases studied and that the differences are usually less than the data statistical uncertainties.

\begin{figure}[t]
    \centering
    \includegraphics[scale=0.75]{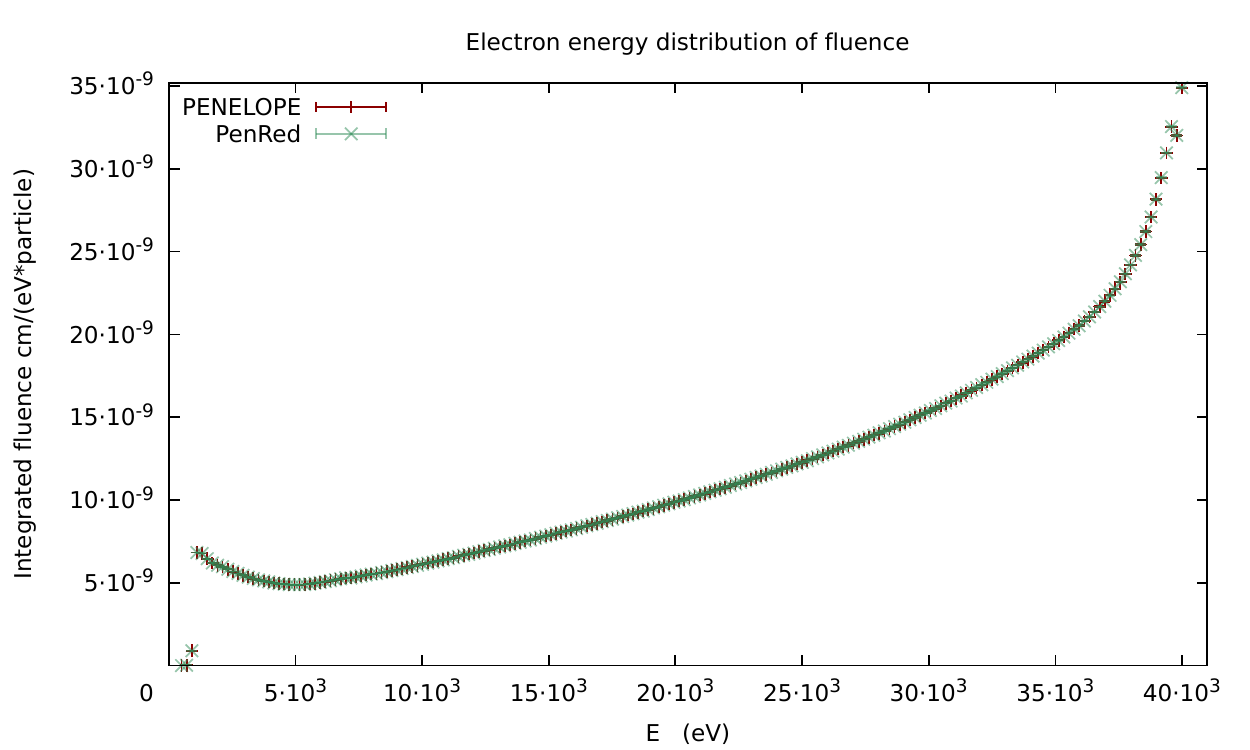}
    \caption{Simulation results of the electron fluence energy distribution integrated over the detector volume of example {\tt 1-disc} with variance reduction techniques. The red (dark) colour represents results from the original PENELOPE code, while green (light) points are the results from PenRed. Some transparency to PenRed points is applied because both curves overlap. Error bars enclosure a deviation of $2\sigma$.}
    \label{fig:ex1fluence}
\end{figure}

For the sake of clarity and easy of reading, only the results for the example {\tt 1-disc-vr} will be presented. Its setup consists of a homogeneous copper cylinder whose radius and height can be set by the user in the geometry file {\tt disc.geo}. The source is a point-like gun that produces a uniform conical beam of electrons with initial energy $40$ keV in the direction of the $Z$ axis and a narrow semi-aperture of $5$ degrees, as specified in the input file {\tt disc.in}. The electrons impact on the cylinder from below. This example has two variants: with and without variance reduction (VR) techniques. The results of the former will be presented below so as to check the PenRed implementation of VR methods. Interaction forcing techniques used in this example increase both the electron bremmstrahlung emission and hard inelastic collision probabilities by a factor of $2000$ and $200$, respectively. In addition, a splitting factor of $2$ on bremmstrahlung and x-ray produced photons was applied. The simulation parameters were $E_{abs} = 1$ keV, $C_{1} = C_{2} = 0.05$ and $W_{cc} = W_{cr} = 1$ keV. The cylinder itself is defined as an energy-deposition, impact and angular detector. The {\tt pen\_main} program was run for about $10.9$ hours to generate $8\times 10^6$ histories, which corresponds to a simulation speed of about $204$ histories per second. Figure \ref{fig:ex1fluence} displays a comparison between the results from PENELOPE and PenRed of the electron energy distribution of fluence integrated over the detector volume.

\begin{figure}[htp]
    \centering
    \includegraphics[scale=0.75]{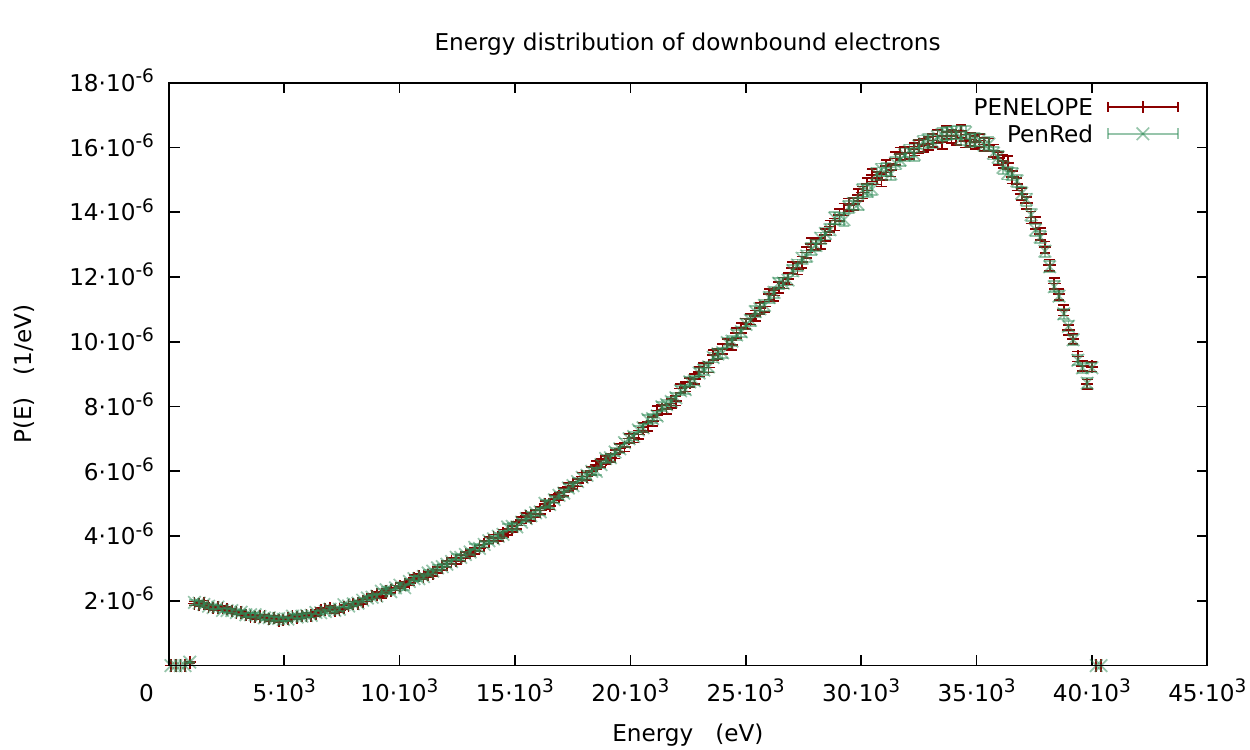}
    \caption{Downbound probability energy distribution for electrons simulated in the example {\tt 1-disc} using VR techniques. The red (dark) colour  represents results from the original PENELOPE code simulations, while green (light) dots are the results from PenRed. For the sake of clarity, some transparency to PenRed points is applied because both curves overlap. Error bars enclosure a deviation of $2\sigma$.}
    \label{fig:ex1downbound}
\end{figure}

In addition, figure \ref{fig:ex1downbound} displays another relevant quantity simulated in this example, namely the probability energy distribution of downbound electrons, i.e. those that escape from the material system in the negative $Z$ direction. 

As can be seen from figs. \ref{fig:ex1fluence} and \ref{fig:ex1downbound}, the results from PENELOPE and PenRed are perfectly compatible within statistical error bars. Moreover, the results are bin-by-bin equivalent with differences that are smaller than statistical errors.

\subsection{Multi-threading tests}
\label{sec:multithreading}

\begin{figure}[t]
    \centering
    \includegraphics[scale=0.75]{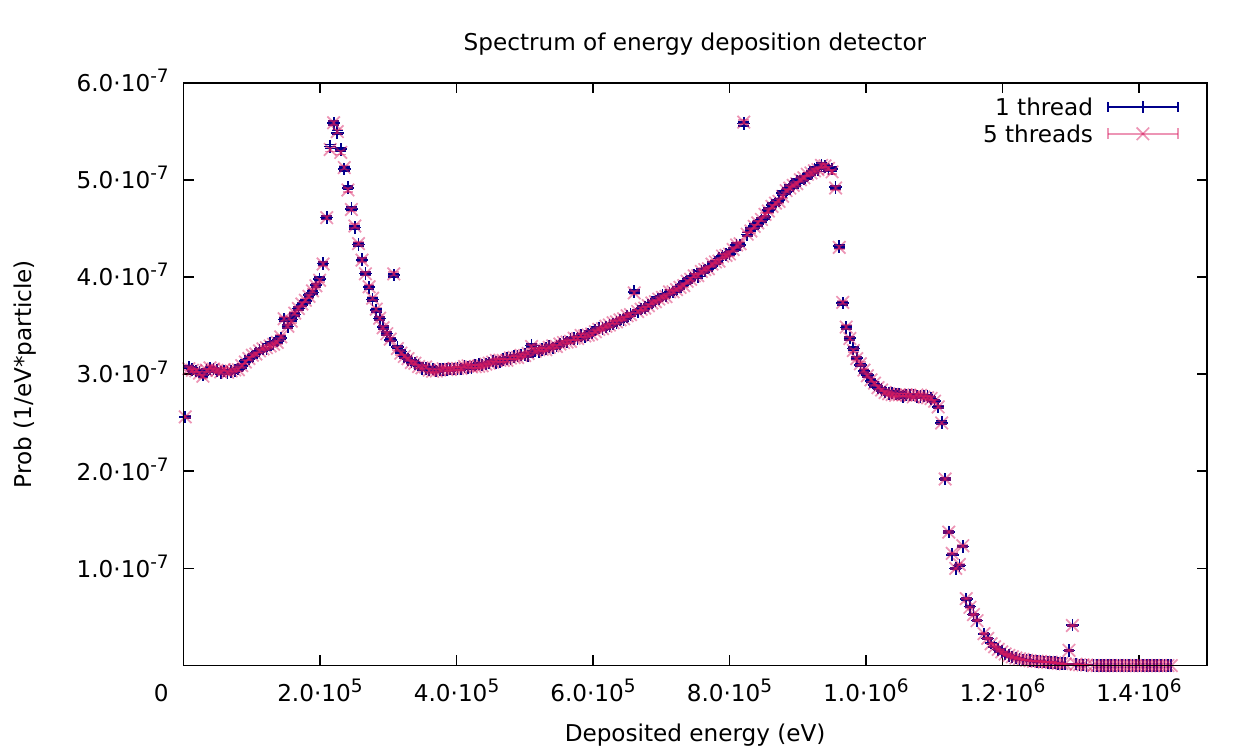}
    \caption{Energy deposition spectrum in a $Na$ scintillator simulation (example {\tt 3-detector}). The colour blue (dark) represents results from single thread simulations, while purple (light) dots are the results from runs on $5$ threads. For the sake of clarity, some transparency to the points corresponding to the runs on $5$ threads is applied because both curves overlap. Error bars enclosure a deviation of $2\sigma$.}
    \label{fig:ex3EndDetMulti}
\end{figure}

In this section, the validation of the multi-threading operation of PenRed is discussed. The same tests with the same simulation parameters, computer setup and compiler options as those done in section \ref{sec:PENELOPEtests}, were carried out by running the PenRed {\tt pen\_main} program on a single thread and on $5$ threads so as to compare the results. The data of each pair of the corresponding output files was compared bin by bin against each other. Figure \ref{fig:ex3EndDetMulti} shows results of the energy deposition spectrum for the PENELOPE example {\tt 3-detector}.

The material system of this example consists of a cylindrical $NaI$ scintillator detector with thin $Fe$ backing. A point-like Co-60 gamma-ray source emits a photon pencil beam in the $-Z$ direction with equiprobable energies $1.17$ and $1.33$ MeV. The photons impinge on the $NaI$ crystal from above. No VR is applied in this example.

All comparisons carried out showed that the results from a single thread and multiple thread simulations are statistically compatible. Therefore, we can ensure that our multi-threading program works properly. In order to reproduce our results, the user only needs to change the parameter $nthreads$ in the provided input configuration files of any of the examples.

\subsection{DICOM tests}
\label{sec:DICOMtest}

In order to validate the PenRed DICOM capabilities, a GATE \cite{Jan_2011} internal targeted radionuclide therapy (TRT) example has been reproduced \footnote{https://davidsarrut.pages.in2p3.fr/gate-exercices-site/docs/exercice5/}. It consists of a treatment in which $^{90}Y$ radionuclides are administrated to a patient to irradiate a liver injury. The material and density information is extracted from a patient CT image, while its associated emitting source spatial distribution is provided as a voxelized source generated using a patient SPECT activity map. Since the GATE example uses the MHD format for input images, a Python script using ITK \cite {10.3389/fninf.2013.00045} has been implemented to convert the images to the DICOM format, which can be directly processed by PenRed.

A voxel-wise comparison between the results from PenRed and Gate has been carried out following the methodology described in \cite{doi:10.1118/1.4953198}. In this work, the statistical variable z-score is defined as follows,

\begin{equation}
    z(\vec{r}) = \frac{D^{PenRed}(\vec{r})-D^{GATE}(\vec{r})}{D^{GATE}(\vec{r}) \times \sigma_{tot}}
\end{equation}

\noindent where $D^{PenRed}(\vec{r})$ and $D^{GATE}(\vec{r})$ are the absorbed doses at voxel $\vec{r}$ calculated by PenRed and GATE, respectively, and $\sigma_{tot}$ is the standard deviation of the distribution $(D^{PenRed}(\vec{r})-D^{GATE}(\vec{r}))/D^{GATE}(\vec{r})$. The resulting z-scores are compared to a standard normal distribution in the form of a normalised frequency histogram, which directly indicates the distribution of dose differences spanning all voxels.

As many voxels present a zero or zero compatible energy deposition, by taking them into account would produce a wide distribution with a spike centred at zero, which would hide the discrepancies from voxels with a significant contribution to the total deposited dose. For this reason, the comparison has been made by considering only the voxels with a registered deposited dose of, at least, $0.1 \%$ of the maximum deposition. As the source is distributed throughout the body of the patient, it includes all the regions of interest.

\begin{figure}[h]
    \centering
    \includegraphics[scale=0.6]{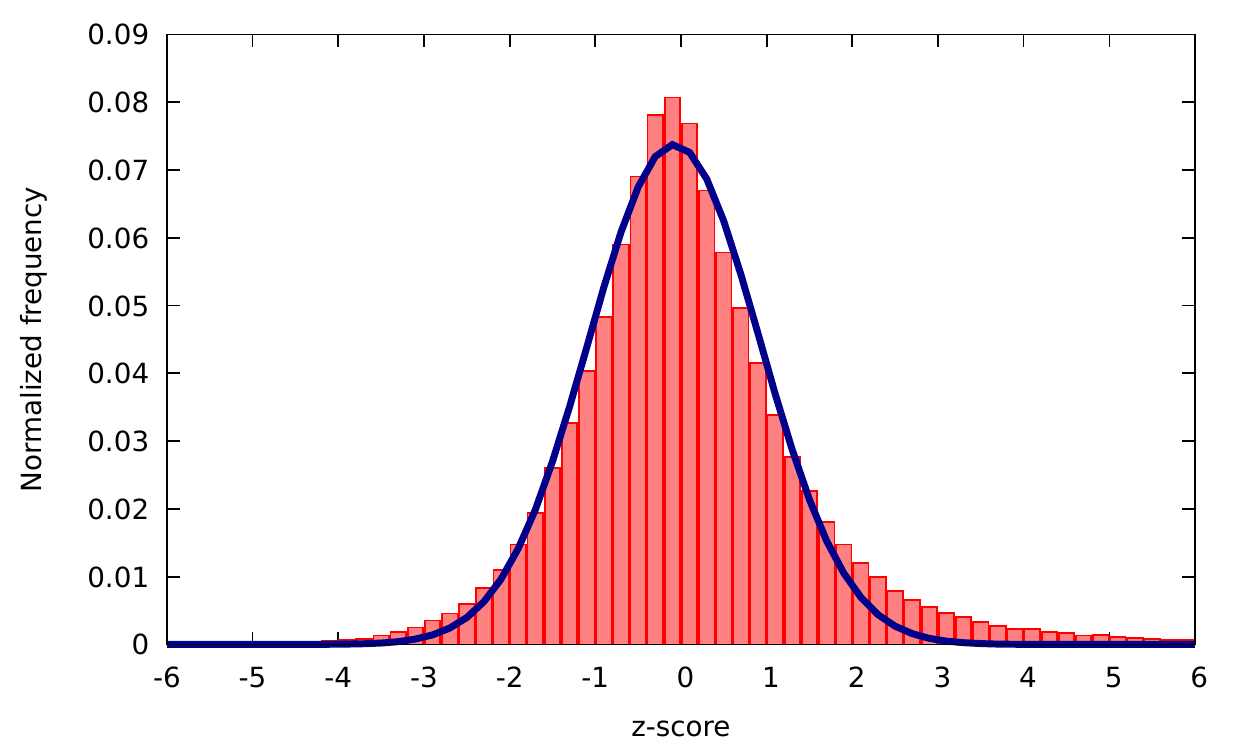}
    \caption{Frequency histogram of z-score values for voxels with a registered dose of, at least, $0.1\%$ of the maximum deposition. The dark line represents a Gaussian fit to the histogram with mean $\mu=-0.078$ and $\sigma=1.002$.}
    \label{fig:DICOMcomp}
\end{figure}

\begin{figure}[h]
    \centering
    \includegraphics{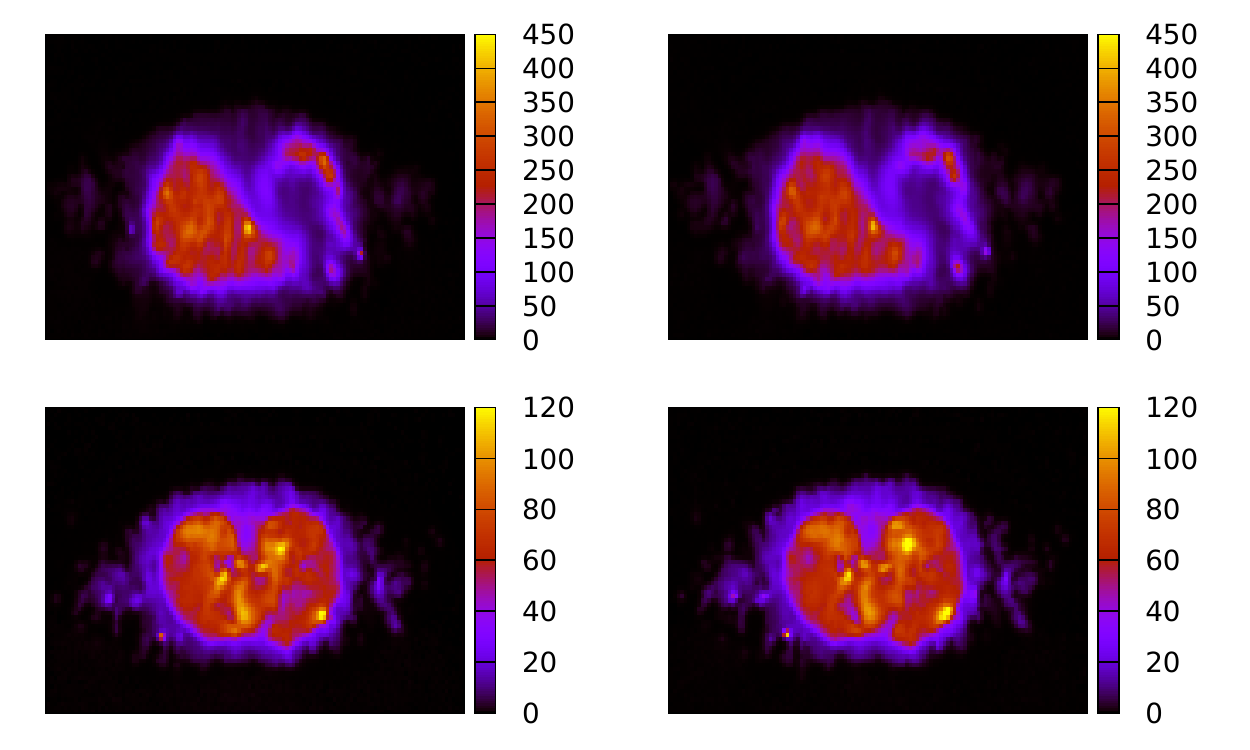}
    \caption{Dose distribution in $eV/g$ per history for the TRT example discussed in the text on the z-planes $127$ (top) and $156$ (bottom). The left panels correspond to results obtained by PenRed, while the right ones has been generated with GATE.}
    \label{fig:DICOMmapdose}
\end{figure}

The results for the frequency distribution of z-scores compared to a Gaussian distribution are displayed in figure \ref{fig:DICOMcomp}. They indicate that the z-score distribution is close to a normal one with a mean near zero ($\mu=-0.078$) and a standard deviation of approximately $1$ ($\sigma=1.002$). However, some non statistical differences can also be observed between both distributions. These discrepancies could be due to the variations between Gate and PenRed in the implementation of the physics models, transport algorithms and the material database. Indeed, discrepancies between different Monte Carlo codes have been reported in the literature, specially when electron transport is involved \cite{Archambault_2015}, \cite{doi:10.1089/cbr.2008.0573}, \cite{doi:10.1118/1.1637970}, \cite{pub.1117066165}.

Finally, figure \ref{fig:DICOMmapdose} shows the dose distributions for two $Z$ planes generated by PenRed (left) and GATE (right). As can be seen, they agree well within expected statistical and systematic uncertainties. This conclusion is also valid for all planes. Therefore, considering the differences between Gate and PenRed implementations mentioned above, the comparison of dose distributions together with the z-score frequency histogram, indicate that PenRed is capable to properly process DICOM images and extended voxelized sources.

\section{Results}
\label{sec:results}

In this section, the parallel capabilities of PenRed are discussed and tested. Firstly, a comparison of the scalability of PenRed in multi-threaded executions with both the PENELOPE main program and the PenEasy code is performed. Secondly, the behaviour of PenRed speed-up is studied. Finally, in order to maximise the efficiency of the simulations with PenRed, a performance test is carried out by using two different compilers.

\subsection{Scalability comparison}
\label{sec:paralBehav}

As discussed in section \ref{sec:parallelism}, the communication among threads or processes is only required at the end of the simulation and represents a negligible amount of time. Thus, MPI executions on different nodes, i.e. running on distributed memory architectures, show a perfect scalability. However, in shared memory parallel executions, the different threads or processes compete for the same resources, which affects the simulation speed-up and scalability. Therefore, in this section we will study the efficiency of parallel simulations running PenRed on a shared memory system.

In order to evaluate the scalability and performance of our implementation, a comparison between PenRed and the 2019 versions of the PENELOPE and PenEasy main programs has been carried out. As the simulation speed depends strongly on specific characteristics such as geometry or the tallies used, we will first compare the capabilities of the different main programs involving only the kernel components that are common to every simulation. To avoid the performance overhead of using specific configurations, such as tallies or complex geometry characteristics, a very simple geometry is simulated. It consists of a single plane dividing the space into two parts, a void and a water region. The source is a monoenergetic point-like one located in the void region and the emitted beam is directed towards the water region perpendicularly to the plane.

Moreover, to isolate as much as possible the kernel components, we perform three different tests using the configuration above. In the first one, the source emits electrons with energy $40$ keV and absorption energies of $1$ keV for electrons and ``infinite'' for photons, i.e. the tracking of photons is disabled. In the second test, photons of $1$ MeV and absorption energy of $1$ keV are emitted, while electron tracking is disabled. Finally, the third test has the same characteristics as the second one but electron tracking is enabled up to $100$ keV. 

To achieve parallelism with the three codes, two different approaches have been employed. In the case of PenRed, the builtin multithreading support has been utilised. However, for the PENELOPE and PenEasy main programs, a bash script to launch multiple processes in parallel has been used, as in \cite{BADAL2006440}.

All the simulations of the tests have been performed with the very same material and geometry files for the three main programs. In order to measure their capabilities to scale up, each simulation has been repeated with increasing number of processes, for PENELOPE and PenEasy, or threads, in the case of PenRed. Indeed, to scale the complexity of the execution, the total number of histories to be simulated has been scaled with the number of concurrent processes or threads. For example, if four processes are to be executed, the total number of histories is four times the number of histories used for the single process test. 

The tests have been run on a single node with an AMD Ryzen 7 2700 processor, $16$ GB RAM and Fedora $30$ as OS. This processor has $8$ physical cores with a total of $16$ logic threads. The two FORTRAN codes have been compiled with the GNU Fortran (GCC) version 9.3.1, while PenRed has been complied with the gcc C++ version 9.3.1. All codes have been compiled with the -O2 and -march=native optimisation flags.

\begin{figure}[!htb]
    \centering
    \includegraphics[scale=1.0]{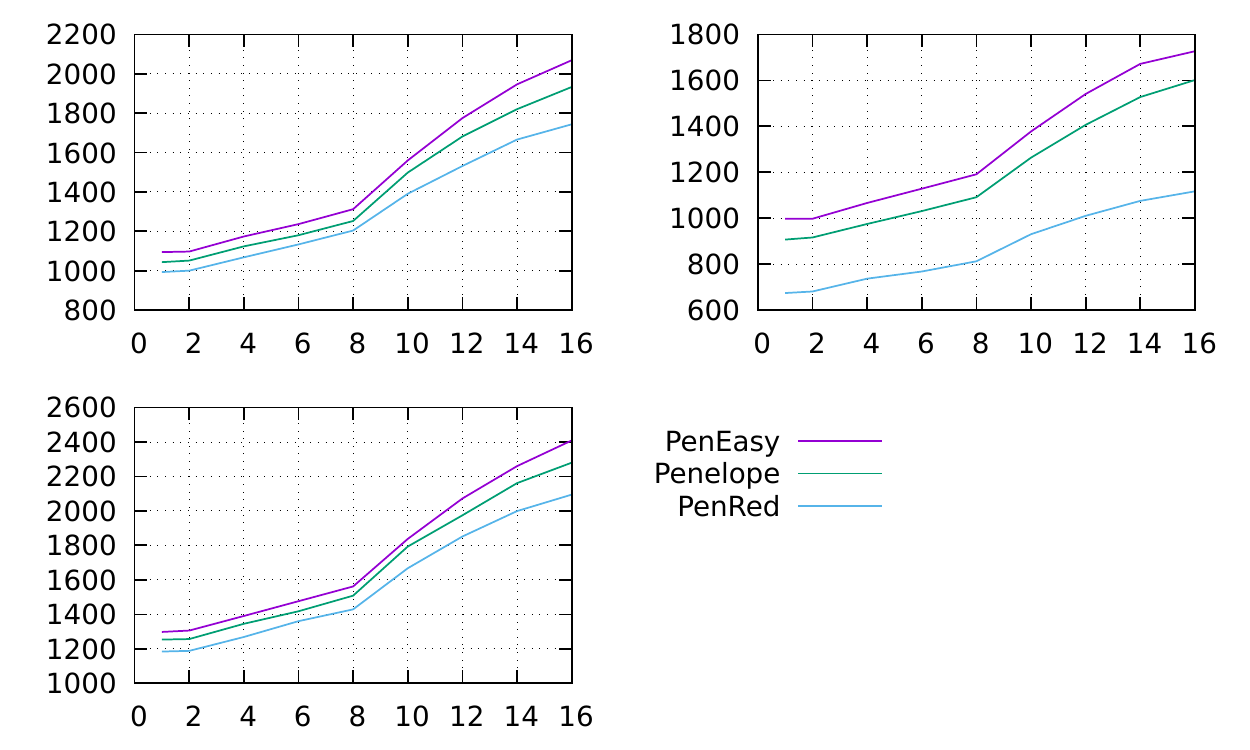}
    \caption{Execution times in seconds for the electron-only (upper left), photon-only (upper right) and combined (lower left) simulation tests as a function of the number of processes/threads.}
    \label{fig:timesScalab}
\end{figure}

\begin{figure}[!htb]
    \centering
    \includegraphics[scale=1.0]{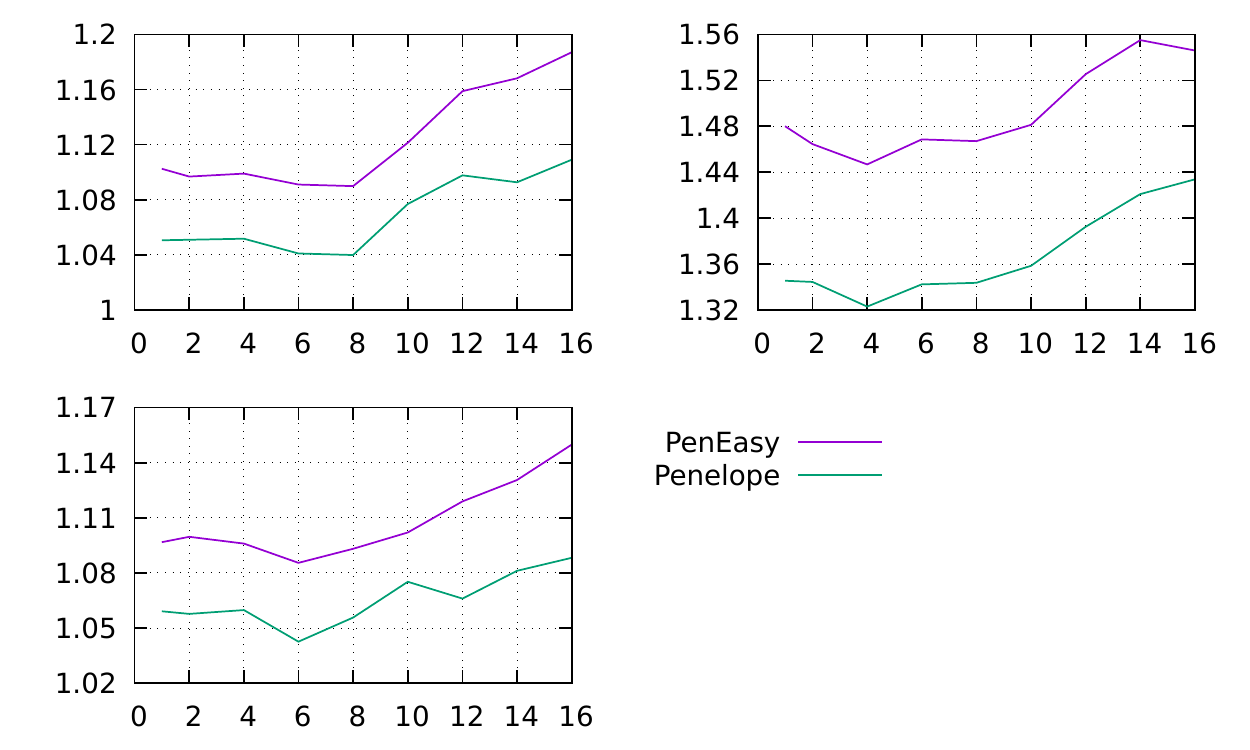}
    \caption{Execution times relative to PenRed for the electron-only (upper left), photon-only (upper right) and combined (lower left) simulation tests as a function of the number of processes/threads.}
    \label{fig:compScalab}
\end{figure}

The test results are shown in figures \ref{fig:timesScalab} and \ref{fig:compScalab}. The former displays the total execution time as a function of the number of processes or threads. The latter presents the same quantity but relative to PenRed execution times. As can be seen, these results suggest that the PenRed implementation of kernel components is more efficient, specially on photon transport. Furthermore, in all tests PenRed provides a better scalability when the number of processes/threads is greater than the number of physical cores. Moreover, in this region, the execution times of the Fortran codes increase with respect to those of PenRed.

To complete the scalability comparison study, a test based on the variant without VR of the PENELOPE example described in section \ref{sec:verification}, {\tt 1-disc-novr}, has been carried out. However, being an example of the PENELOPE package, the PenEasy main program does not implement all the involved tallies, which should slightly affect the total execution time. On the contrary, PenRed implements all the required tallies. Notice that the choice of a non VR version of the example is due to the fact that PenEasy and PENELOPE main programs implement different types of VR.

\begin{figure}[h]
    \centering
    \includegraphics[scale=0.75]{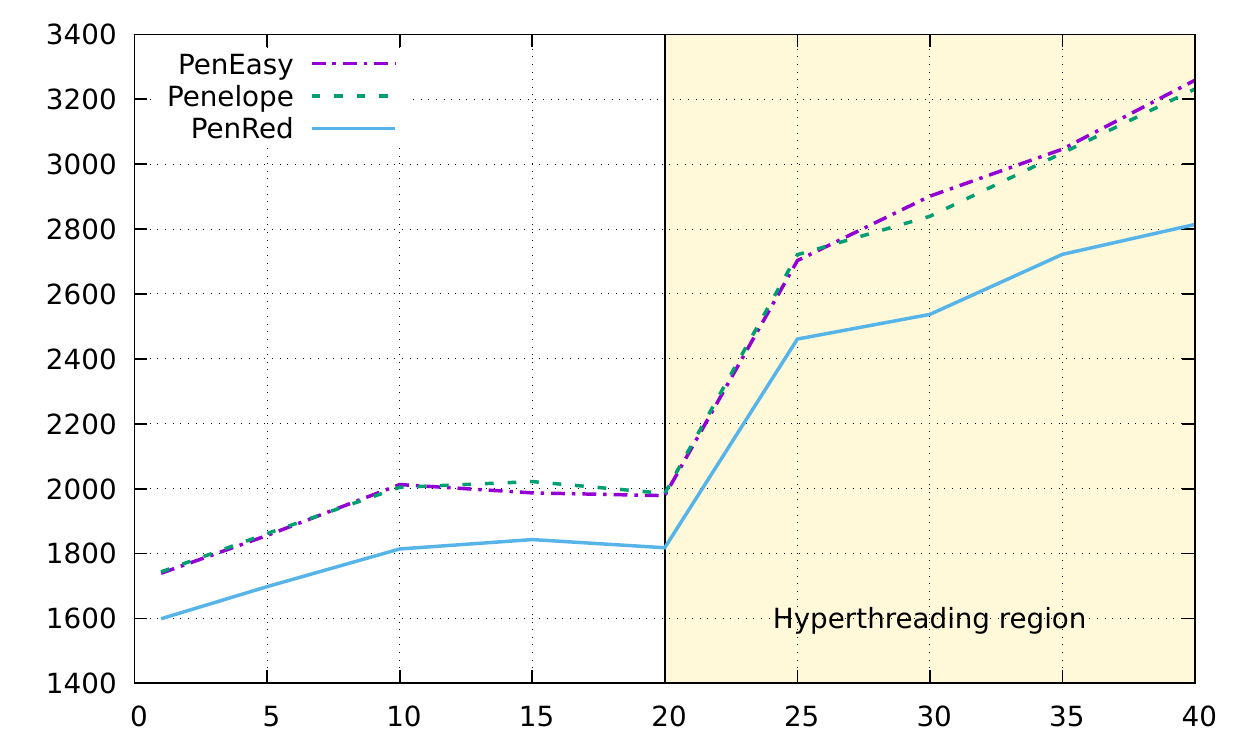}
    \caption{Comparison of the execution times in seconds for the three codes as a function of the number of processes/threads used.}
    \label{fig:timesDiscHydra}
\end{figure}

This last test has been run on the same node used for the code validation (see section \ref{sec:verification}), which consists of two processors with a total of $40$ logical threads. In this case, the number of simulation processes or threads has been increased by a factor of 5 in each iteration. The resulting simulation times are shown in figure \ref{fig:timesDiscHydra}, where differences of up to $15\%$ between PenRed and both, Penelope and PenEasy main programs, can be noted. Furthermore, the behaviour observed in this test is analogous to that of the previous tests, namely PenRed takes the lead in execution time as the number of threads increases exceeding the number of physical cores.

Thus, the suitability of using a multithreading approach rather than multiprocess executions on shared memory architectures is demonstrated, specially on high parallel processors. 

\subsection{Speed-up behaviour}
\label{sc:speedup}

In this section, tests focus on the speed-up behaviour of the PenRed code are described. Instead of scaling the problem complexity with the number of threads, in this analysis the former remains constant as the latter increases. To measure the speed-up, the quantity $S_n$ is used. It is defined as
\begin{equation}
    S_n = \frac{time_1}{time_n}
    \label{eq:speed-up}
\end{equation}

\noindent where the subscript $n$ indicates the number of threads, $time_1$ is the simulation time running a single thread and $time_n$ the execution time of the same simulation using $n$ threads. Ideally, $S_n$ should tend to the number of threads $n$, which is the maximum speed-up that can be achieved. 

\begin{figure}[!htb]
    \centering
    \includegraphics[scale=0.75]{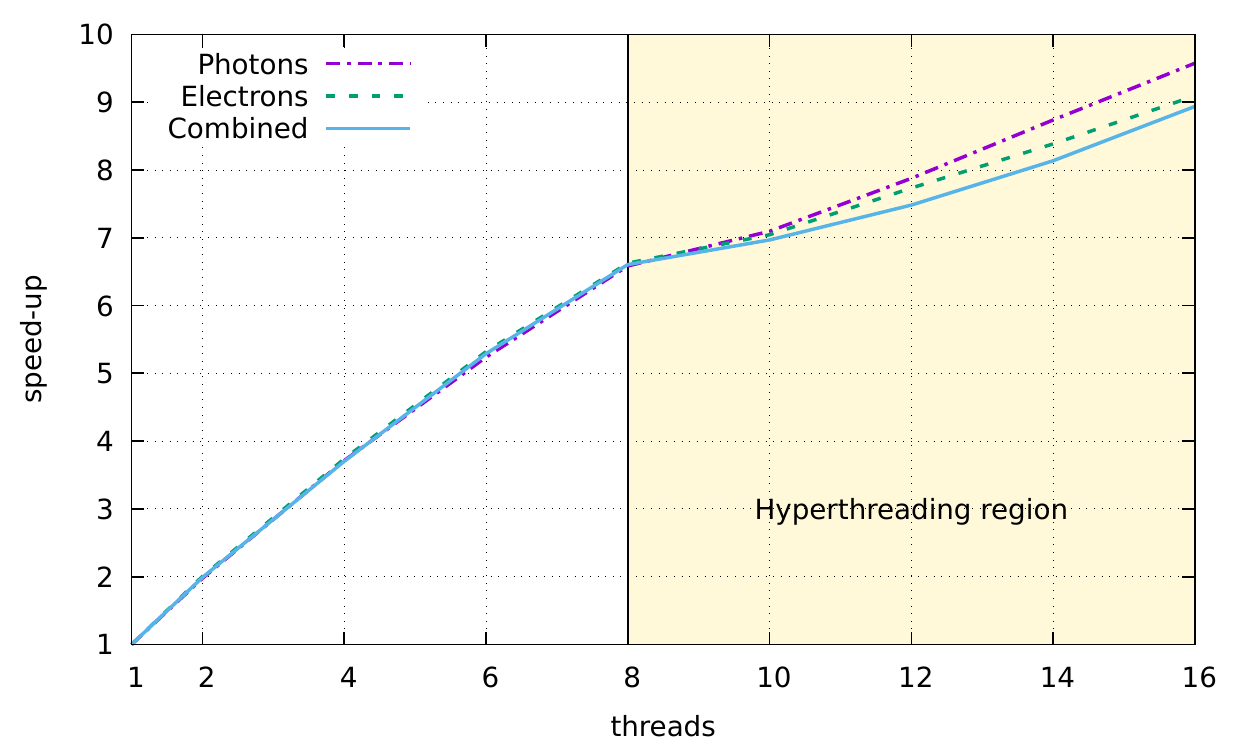}
    \caption{PenRed speed-up values of the same tests as in the scalability analysis (figures \ref{fig:timesScalab} and \ref{fig:compScalab}). The figure is divided into two zones. The left one, from $1$ to $8$ threads, corresponds to executions with one thread per core. The right zone, from $9$ to $16$ threads, corresponds to the situation where threads share physical cores.}
    \label{fig:speedUpSimple}
\end{figure}

Firstly, the speed-up for the same three tests used in the scalability analysis has been measured. The results are shown in figure \ref{fig:speedUpSimple}. In all cases, a linear speed-up has been obtained until the number of threads exceeds the number of physical cores. A linear fit to the speed-up data in this region gives a slope of about $0.86$ for all three cases. After this point, logical threads share the available processor cores. As can be seen, the photon-only simulation shows a better speed-up than both, electron-only and combined simulations. Nevertheless, in all cases it is worth using all the processor logical threads as the simulation time is reduced by $45\%$, $37\%$ and $35\%$ for the photon-only, electron-only and combined tests respectively, compared to the execution times with only $8$ threads.

\begin{figure}[!htb]
    \centering
    \includegraphics[scale=0.75]{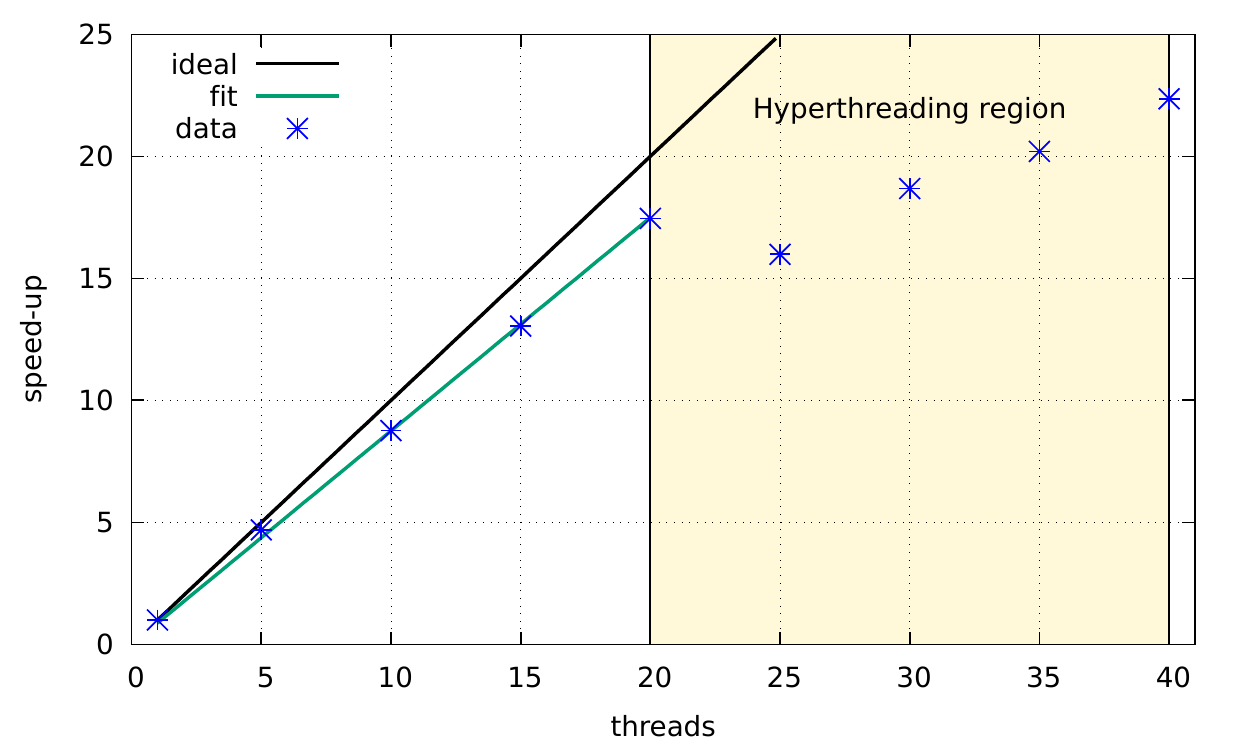}
    \caption{Speed-up values of the PENELOPE example {\tt 1-disc-novr}. The figure is divided into two zones. The left one, from $1$ to $20$ threads, corresponds to executions with no hyperthreading usage. The right zone, from $21$ to $40$ threads, corresponds to the hyperthreading region. The black line represents a perfect ideal speed-up while the green line is a linear fit to the speed-up data from $20$ threads or less.}
    \label{fig:speed-upHydraDisc}
\end{figure}

In addition, the speed-up of the same examples as in section \ref{sec:PENELOPEtests} has been measured using the node with $40$ logic threads. We found that all the tested examples show a similar speed-up behaviour. In figure \ref{fig:speed-upHydraDisc}, the values of $S_n$ for the example {\tt 1-disc-novr} as a function of the number of threads are presented. The green line represents a linear fit to the measured speed-up data when a maximum of one thread was executed on each physical processor core, corresponding to the zone between $1$ and $20$ threads. In this region, as discussed above, a linear correlation between the speed-up and the number of running threads was obtained. The value of the slope was $0.87$. Again, it is worth using the full hyper-threading capabilities ($40$ threads) as a $28\%$ reduction on the total simulation time compared to that obtained employing only $20$ threads is achieved.

Thus, the speed-up tests described above demonstrate that PenRed is well suited for massively parallel infrastructures, as takes efficiently advantage of all the available resources.

\subsection{Compiler tests}
\label{sec:compilerPerformance}

To maximise the program efficiency, the speed-up tests of all PENELOPE examples were repeated using the Intel C++ compiler \cite{IntelC++} ({\tt icc}) version 19.0.5.281 on the Intel processor node described in section \ref{sec:verification}. We found that the speed-up behaviour is very similar for both compilers. However, tests performed by running on a single thread show that the code compiled with {\tt icc} is faster (by about $50$-$60\%$). These results are partially shown in Table \ref{tb:gccvsintel}, where the simulation speed of three PENELOPE examples are collected.

\begin{table}[H]
\centering
\begin{tabular}{|c|c|c|c|}
\hline    Example & GCC & icc & Increase ($\%$)\\
\hline    1-disc-vr & $2.037\cdot 10^2$ & $3.039\cdot 10^2$ & $49.16$ \\
\hline    3-detector & $2.623\cdot 10^3$ & $4.167\cdot 10^3$ & $58.88$ \\
\hline    5-accelerator & $2.097\cdot 10^2$ & $3.194\cdot 10^2$ & $52.31$ \\
\hline
\end{tabular}
\caption{Comparison between simulation speeds (histories/s) using the very same code compiled with GNU {\tt g++} version $7.3.1$ and Intel {\tt icc} version 19.0.5.281. Both compilations were run on the same computer setup with a single thread. The first column specifies the example tested. The second and third columns contain the simulation speed in histories per second for the code compiled with {\tt g++} and {\tt icc}, respectively. Finally, the last column shows the increase in the simulation speed in percentage.}
\label{tb:gccvsintel}
\end{table}

\section{Conclusions and future work}
\label{sec:conclusions}

PenRed provides a flexible, object-oriented, parallel and general purpose framework for Monte Carlo simulations of radiation transport in matter based on PENELOPE physics. As it has been verified, these features have been achieved without compromising simulation speed. Furthermore, the efficiency has been improved significantly. PenRed includes all the PENELOPE physics models. It has been thoroughly tested that PenRed reproduces the cross-sections of PENELOPE and that, within statistical errors, gives also the same results for the complete set of PENELOPE examples. Moreover, our results can be reproduced by the user using the input files provided in the PenRed package. In addition, our code system includes support to simulate voxel geometries and DICOM images, which make it suitable for performing Monte Carlo simulations of medical treatments.

Regarding parallelism, PenRed incorporates a mixed model for handling both distributed and shared memory parallelism via MPI and the standard C++ thread library, respectively. We have shown that PenRed achieves good speed-ups for both types of parallelism, which denotes that these features have been efficiently implemented. Furthermore, a builtin load balance system has been implemented to optimise executions on heterogeneous environments.

On the other hand, the modular structure of PenRed allows the developers to incorporate new custom components (particle sources, geometries, tallies, etc.) which can be included with no modification to the original PenRed code. The new components will be automatically usable in parallel computations without any previous knowledge of parallel programming.

In future versions of PenRed, we will implement new tallies and source models. In addition, we plane to improve the parallelism implementation to accelerate the execution time and adapt it to perform generic simulations on specific hardware accelerators, such as GPGPUs and FPGAs.

\section*{Acknowledgements}

The authors are deeply indebted to F. Salvat for many comments and suggestions, for clarifying many subtleties of the simulation algorithms of the transport of particles through matter, specially using PENELOPE, and for his patience and understanding. The work of V. Gimenez-Alventosa was supported by the program ``Ayudas para la contratación de personal investigador en formación de carácter predoctoral, programa VALi+d'' under grant number ACIF/2018/148 from the Conselleria d’Educació of the Generalitat Valenciana and the ``Fondo Social Europeo'' (FSE). V. Gimenez acknowledges partial support from FEDER/MCIyU-AEI under grant FPA2017-84543-P, by the Severo Ochoa Excellence Program under grant SEV-2014-0398 and by Generalitat Valenciana through the project PROMETEO/2019/087.

\bibliographystyle{elsarticle-num}

\bibliography{penred}

\end{document}